# A comparison of two methods for detecting abrupt changes in the variance of climatic time series

Sergei N. Rodionov[1]

[1]Climate Logic LLC, Superior, CO, 80027, U.S.A.

*Correspondence to*: Sergei N. Rodionov (rodionov.sergei@gmail.com)

**Abstract.** Two methods for detecting abrupt shifts in the variance – Integrated Cumulative Sum of Squares (ICSS) and Sequential Regime Shift Detector (SRSD) – have been compared on both synthetic and observed time series. In Monte Carlo experiments, SRSD outperformed ICSS in the overwhelming majority of the modelled scenarios with different sequences of variance regimes. The SRSD advantage was particularly apparent in the case of outliers in the series. When tested on climatic time series, in most cases both methods detected the same change points in the longer series (252-787 monthly values). The only exception was the Arctic Ocean SST series, when ICSS found one extra change point that appeared to be spurious. As for the shorter time series (66-136 yearly values), ICSS failed to detect any change points even when the variance doubled or tripled from one regime to another. For these time series, SRSD is recommended. Interestingly, all the climatic time series tested, from the Arctic to the Tropics, had one thing in common: the last shift detected in each of these series was toward a high-variance regime. This is consistent with other findings of increased climate variability in recent decades.

## 1 Introduction

A concept of regime shifts, i.e. abrupt structural changes in climatic time series, has gained popularity in recent decades. Literature reviews (Beaulieu et al., 2012; Liu et al., 2016; Rodionov, 2005a) show that most of the effort in this area is directed towards changes in the mean, where numerous methods of shift (or change point) detection have been developed. Changes in climate variability has received less attention, both in documenting those changes and developing of methods for their detection. In many applications, however, the potential impact of changes in climate variability may be as great or greater than the impact of changes in climate means (Hansen et al., 2012; Katz, 1988).

Patterns of climatic variability in the first half of the twentieth century were investigated by Schuurmans (1984), who showed that the minimum interannual variability of surface air temperature in western Europe occurred during the 1890-1920 period. In central Europe, this minimum was less pronounced and shifted to a later period. For the Northern Hemisphere as a whole, the minimum frequency of temperature extremes occurred in the decade of 1920-1929.

More recently, Scherrer et al. (2005) investigated standardized distribution changes for seasonal mean temperature in central Europe and found that temperature variability showed a weak increase (decrease) in summer (winter) from 1961 to 2004, but these changes were not statistically significant at the 10% level. Peel and McMahon (2006) reported that the interannual variability of temperature and precipitation marginally decreased since 1970. Coherent regions of decreasing interdecadal temperature variability (after 1970) were observed in Japan, East Coast USA, Pacific Northwest USA, western Canada, north-western Russia and eastern Australia. Huntingford et al. (2013) demonstrated that although fluctuations in annual temperature showed substantial geographical variation over the past few decades, the time-evolving standard deviation of globally averaged temperature anomalies was stable. In contrast, Hansen et al. (2012) showed a widening of temperature probability density functions in the past three decades, especially in summer.

Summer extreme events, such as the heat waves of 2003 in western Europe (Chase et al., 2006) and of 2010 in eastern Europe (Barriopedro et al., 2011), sparked debates of whether or not these were signs of global warming, the greatest impact of which could be due to changes in frequency of extreme events, and not just because of a simple increase in the mean. Although the probability of extreme events may be affected by changes in the mean state of the atmosphere (Rhines and Huybers, 2013), changes in the variance maybe just as important. According to Katz and Brown (1992), the more extreme the event, the more important a change in the variability is relative to the mean. Indeed, Schär et al. (2004) calculated that the 2003 heat wave would be extremely unlikely given a change in the mean only. They showed that a recent increase in variability would be able to explain the heat wave.

Still, individual extreme events by themselves are not enough proof that the overall variance is on the rise. Those events may turn out to be just outliers that make the detection of regime shifts in the variance even more difficult. In fact, using very long and homogenized instrumental climatic time series Böhm (2012) found no change in variability during the past 250 years in Europe, not for pressure, not for temperature and not for precipitation. Interestingly, evidence from Greenland ice cores showed that an enhanced year-to-year temperature variability was probably more characteristic of past cold, rather than warm climates (Steffensen et al., 2008). Therefore, in spite of suggestions of higher variability in recent decades (Hansen et al., 2012), there is still considerable uncertainty as to whether it is actually occurring.

Conflicting reports and often statistically inconclusive results underscores difficulties in detecting abrupt changes in the variance of climatic time series. As it will be shown in this paper, it is often hard to detect a regime shift even if the variance increases twofold from one period to another, particularly for short time series, and even more so in the presence of outliers.

The most common statistical tool to study changes in climate variability is the running standard deviation; however, this tool cannot be used in the case of abrupt transitions between variance regimes. Downton and Katz (1993) developed a test to detect and adjust inhomogeneities in the variance of temperature time series. Their test uses a nonparametric bootstrap technique to compute confidence intervals for the discontinuity in variance. More recently, Killick et al. (2010) used a change point analysis to detect abrupt changes in the variance of significant wave heights in the Gulf of Mexico. They adopted a penalized likelihood approach using the Schwarz' information criterion proposed by Yao (1988).

Currently, the most advanced methods of regime shift detection in the variance appear to exist in econometrics, especially quantitative finance, where the concept of stock market volatility is very important. One of the most popular among those methods is the Iterated Cumulative Sum of Squares (ICSS) algorithm developed by Inclan and Tiao (1994). The purpose of this paper is to compare ICSS with the Sequential Regime Shift Detector (SRSD), a method developed by the author (Rodionov 2005b). These two methods are described in Section 2. In Section 3, the methods are tested using Monte Carlo experiments with different sequences of variance regimes, magnitudes of shifts and positions of change points. The effect of outliers is also evaluated. The real-world examples are presented in Section 4. The methods are applied to climatic time series from different geographic zones, from the Arctic to the Tropics. The results are summarized in Section 5.

## 2 Methods

### 2.1 Iterated cumulative sums of squares

Suppose $\{x_i\}, i = 1, \ldots n$ is a series of independent, normally distributed random variables with zero mean and variance $\sigma^2$ that experiences abrupt shifts at unknown time, that is

$$\sigma_i^2 = \begin{cases} \sigma_1^2, & i = 1, \ldots, c_1, \\ \sigma_2^2, & i = c_1 + 1, \ldots c_2, \\ \ldots \\ \sigma_m^2, & i = c_{m-1} + 1, \ldots n. \end{cases} \qquad (1)$$

The task is to find the number of variance regimes $m$ and locations of change points $c_j, j = 1, \ldots, m-1$. To solve this task, Inclan and Tiao (1994) proposed to use the centered (and normalized) cumulative sum (CUSUM) of squares:

$$D_k = \frac{C_k}{C_n} - \frac{k}{n}, \quad k = 1, \ldots, n, \qquad (2)$$

where $C_k = \sum_{i=1}^{k} x_i^2$. The algorithm consists of several steps, dividing time series $\{x_i\}$ into pieces and applying $D_k$ to each of them iteratively.

Step 1. Calculate $D_k(x[i_1:i_2])$ for the entire time series, i.e., $i_1 = 1$ and $i_2 = n$. Let $k^*(x[i_1:i_2])$ be the point at which $max_k |D_k(x[i_1:i_2])|$ is obtained and let

$$M(i_1:i_2) = \max_{i_1 \leq k \leq i_2} \sqrt{(n - i_1 + 1)/2} \, |D_k(x[i_1:i_2])|. \qquad (3)$$

If $M(i_1:i_2)$ is greater than the critical value $D^*$ from Table 1 in Inclan and Tiao (1994), then there is a change point at $k^*(x[i_1:i_2])$ and proceed to step 2a. If $M(i_1:i_2) < D^*$, the is no evidence of variance changes in the series and the algorithm stops.

Step 2a. Evaluate $D_k(x[i_1:i_2])$ for the first part of the series from the beginning up to $i_2 = k^*(x[i_1:i_2])$. If $M(i_1:i_2) > D^*$, step 2a is repeated for a new (smaller) $i_2$, until $M(i_1:i_2) < D^*$. When this occurs, the first point of change is $k_{first} = i_2$.

Step 2b. Now do a similar search for the second part of the series from the change point found in step 1 till the end of the time series, i.e., $i_1 = k^* + 1$ and $i_2 = n$. If $M(i_1:i_2) > D^*$, step 2b is repeated using a new (larger) $i_1$, until $M(i_1:i_2) < D^*$. When this occurs, the last point of change is $k_{last} = i_1 - 1$.

Step 2c. If $k_{first} = k_{last}$, there is just one change point. If $k_{first} < k_{last}$, keep both values as possible change points and repeat steps 1 and 2, but for the middle part of the series, i.e., $i_1 = k_{first} + 1$ and $i_1 = k_{last}$. Each time steps 2a and 2b are repeated, the result can be one or two more change points. Call $m$ the number of variance regimes ($m - 1$ change points) found so far.

Step 3. Check each possible change point $c_j$ by calculating $D_k(x[c_{j-1} + 1: c_{j+1}]), j = 1, ..., m - 1; c_0 = 0, c_m = n$. If $M(c_{j-1} + 1: c_{j+1}) > D^*$, then keep the point; otherwise, eliminate it. Repeat step 3 until the number of change points does not change and the points found in each new pass are "close" (within two observations in our case) to those on the previous pass.

**2.2 Sequential regime shift detector**

The SRSD software package consists of three modules for detection of regime shifts in the mean, variance and correlation coefficient (see www.climatelogic.com). The variance module, which is the focus here, was first described in Rodionov (2005b) and then, as part of a three-step procedure, in Rodionov (2015). Similar to ICSS, SRSD is a CUSUM-type method. An important difference is that ICSS is a retrospective algorithm, whereas SRSD employs a sequential approach. While ICSS can be applied only after all the data is collected, SRSD treats the data sequentially, one data point at a time. This allows using SRSD for monitoring of regime shifts beyond the history period $1, ..., n$.

The SRSD algorithm is based on the $F$-test that compares the ratio of sample variance for the "current" regime, $s_j^2$ to that for the "new" regime to be detected, $s_{j+1}^2$, with the critical value $F_{cr}$:

$$F = \frac{s_j^2}{s_{j+1}^2} \lessgtr F_{cr}. \qquad (4)$$

Here $F_{cr}$ is the value of the $F$-distribution with $\nu_1$ and $\nu_2$ degrees of freedom and a significance level $p$ (two-tailed test), $F_{cr} = F(p/2, \nu_1, \nu_2)$. The degrees of freedom are calculated as $\nu_1 = \nu_2 = l - 1$, where $l$ is called the cut-off length, a parameter that allows to focus on certain time scales of variance regimes. More about the effect of $p$ and $l$ on regime shift detection in the next section.

For the new regime to be statistically different from the current regime, variance $s_{j+1}^2$ should be greater than critical variance $s_{cr}^{2\uparrow}$, if the variance increases, or smaller than $s_{cr}^{2\downarrow}$, if the variance decreases, where

$$s_{cr}^2 = \begin{cases} s_{cr}^{2\uparrow} = s_j^2 F_{cr}, \\ s_{cr}^{2\downarrow} = s_j^2 / F_{cr}. \end{cases} \qquad (5)$$

When a new observation $x_i$ arrives at time $i$, it is checked against $s_{cr}^2$. If $x_i^2$ falls outside the interval $[s_{cr}^{2\downarrow}, s_{cr}^{2\uparrow}]$, this point in time is marked as a potential change point $c_p$, and the subsequent $l - 1$ data points are used to test the null hypothesis of no regime shift. The decision rule is based on the residual sum of squares index (RSSI):

$$RSSI = \frac{1}{l} \sum_{i=c_p}^{k} (x_i^2 - s_{cr}^2), k = c_p, c_p + 1, ..., c_p + l - 1. \qquad (6)$$

If during this testing period RSSI remains positive in the case of increasing variance, or negative in the case of decreasing variance, the null hypothesis of a constant variance is rejected, and $c_p$ becomes a true change point $c_{j+1}$. If RSSI changes its sign, the test for $c_p$ stops, because the null hypothesis cannot be rejected. The observation $x_i$ is included in the current regime $j$, and the regime variance $s_j^2$ is recalculated incrementally as in Finch (2009):

$$s_{new}^2 = \frac{n_j s_{old}^2 + x_i^2}{n_j + 1}, \tag{7}$$

where $n_j$ is the length of regime $j$ before adding a new observation.

**2.2.1 Handling outliers**

An outlier is a data point whose value is substantially different from the other data points in a sample. The presence of outliers can lead to large errors in estimates of regime statistics and greatly affect the timing of regime shifts. This is particularly true for the variance, because the data values are squared.

When dealing with outliers, it is desirable to leave an observation intact if it falls within a "normal" range of variation, and assign it a small weight if it is outside that range (Huber, 1981). In SRSD, each observation $x_i$ in the current regime $j$ is assigned a Huber-type weight $w_i$, which is defined as

$$w_i = min\left(1, h\frac{scale}{|x_i|}\right), \tag{8}$$

where $h$ is a tuning constant and *scale* is first estimated as the median absolute deviation (MAD)

$$MAD = median(|x_i|), \quad i = c_j, c_j + 1, \dots, c_p - 1. \tag{9}$$

After the weights are estimated, the regime variance is calculated as

$$s_j^2 = \frac{\sum_{i=c_j}^{c_p-1} w_i^2 x_i^2}{V_1 - (V_2/V_1)}, \tag{10}$$

where

$$V_1 = \sum_{i=c_j}^{c_p-1} w_i^2, \qquad V_2 = \sum_{i=c_j}^{c_p-1} w_i^4. \tag{11}$$

To improve accuracy of the estimates, the weights are recalculated, this time using the weighted standard deviation as *scale*. Then the regime variance is recalculated one more time as in Eq. 10 using more accurate weights.

When a potential regime shift is tested by calculating RSSI, the weights for data points $i = c_p, c_p + 1, \dots, c_p + l - 1$, are assigned as

$$w_i = min\left(1, h\frac{s_j \pm s_{cr}}{|x_i|}\right). \tag{12}$$

If the test cannot reject the null hypothesis and $x_i$ is added to the current regime, the weight for this observation is recalculated using Eq. 8 with $scale = s_j$. The weighted version of Eq. 7 for incremental recalculation of the regime variance is

$$s_{new}^2 = \frac{V_1 s_{old}^2 + w_i^2 x_i^2}{V_1 + w_i^2}. \qquad (13)$$

## 3 Monte Carlo experiments

The majority of experiments were performed using samples of size $n = 100$, which is a typical length (in years) of instrumental climatic time series. For each test, 10000 samples of random normally distributed numbers are generated with zero mean and one or two change points in the variance. A change points is defined as in SRSD, that is, as the first point of a regime. Since ICSS defines a change point as the last points of a regime, all the change points detected by ICSS are shifted one step forward. A total of three groups of experiments were conducted: 1) with one change point located in the middle and closer to the ends of time series, 2) with two change points and different combinations of variance regimes, and 3) with outliers. Before starting comparing the two methods, it is important to clarify the role of the significance level $p$ and cut-off length $l$ on regime shift detection by SRSD.

### 3.1 The effect of tuning parameters $p$ and $l$

As shown above, the performance of SRSD can be controlled by three parameters: the significance level $p$, cut-off length $l$, and Huber's tuning constant $h$. This will be denoted as SRSD ($p$, $l$, $h$). Let us clarify the role of the first two parameters. The Huber's tuning constant becomes really important only in the presence of outliers.

The significance level $p$ (also called the target probability level) controls how sensitive SRSD is to shifts in the variance. Since the critical value $F_{cr}$ decreases with the increase of $p$, SRSD becomes sensitive to smaller and smaller shifts in the variance, thus reducing a type II error ("false negative"). On the other hand, a type I error ("false positive") increases, leading to spurious shifts. This is demonstrated in Fig. 1 that shows the frequency of shift detection for different values of $p$. The true change point is located at $i = 51$ and the variance doubles from the first regime to the second. At $p = 0.05$, the percentage of hits (change points detected exactly at $i = 51$) is 4.5. In 57% of the time, SRSD fails to detect any shift in the series (Table 1). As $p$ increases, the percentage of hits also increases, but the tails of the distribution in Fig. 1 become heavier. It is important to note, that since SRSD is a sequential algorithm, the frequency of change points found at the end of the series increases substantially, as illustrated for $p = 0.3$. But these are only *potential* change points that may or may not pass the test. Therefore, when comparing SRSD and ICSS, only the fully resolved change points were considered. More specifically, the numbers of change points detected in each series (as presented in Tables 1-5) were counted for the period $1, ..., n - l$ for both methods.

The choice of $p$ depends on the magnitude of shifts. Figure 2 shows that as the magnitude of a shift increases, the accuracy of shift detection is increasing faster for smaller $p$. Thus, if the ratio of variances between two adjacent regimes is greater than 4, it is better to use $p = 0.05$, which provides better accuracy of detection and fewer false positives, than $p = 0.1$ or $p = 0.3$. If the goal is to detect shifts of smaller magnitude, the choice of $p = 0.1$ or higher would be appropriate. The target probability level $p$ can be considered as the upper limit of the desired significance level of the shifts; the $p$-values, calculated after all shifts in a series have been detected, are usually lower.

The cut-off length $l$ controls the time scale of the detected regimes. As an example, Fig. 3a shows the results for time series with three variance regimes, $\sigma_j^2 = (1, 3, 1)$, and two change points, $c_j = (31, 61)$. The best results were obtained when $l$ was set to be equal the length of the first two regimes ($l = 30$). In 73% of the time, SRSD detected two shifts simultaneously and the percentage of hits was the highest (Table 2). At $l = 20$, the test statistics in Table 2 do not change much, except for the percentage of hits for $c_1$. When the cut-off length is larger than the regime length ($l = 40$), the effect on the test statistics is more dramatic. The percentage of two shifts detected simultaneously in a series sharply decreases, while the percentage of one or zero shifts increases. The first shift is affected more significantly than the second one. For a reversed order of the variance regimes (3, 1, 3), the percentage of hits for both shifts is about the same for all cut-off lengths used (Fig. 3b). Table 2 shows, however, that at $l = 40$, only one of the two shifts is detected most of the time (69%).

In the Monte Carlo experiments below, the values of $p$ are chosen so that to make type I errors about the same for both methods. The values of $l$ are used a bit smaller than a specified regime length, because in practice the exact length of regimes in a series is usually unknown. When working with real time series it is recommended to experiment with different values of $p$ and $l$, in the same way as it is always recommended in spectral analysis to try different window shapes and sizes when smoothing a periodogram.

**3.2 Series with one change point**

The first two experiments were performed with a change point positioned in the middle of the series ($c_1 = 51$), while the variance either increased from one to two (Fig. 4a) or decreased from two to one (Fig. 4b). In both cases, SRSD outperformed ICSS in terms of the correct number of change points detected and percentage of hits (Table 4). Note that ICSS found no shifts at all in about 40% of the generated series.

It is worth noting an asymmetry in the frequency distribution of detected change points, particularly for ICSS. Change points tend to be detected more frequently after $c_1$ than before if the variance increases, and the opposite is true if the variance decreases. This is because of a higher probability density near the mean (zero in this case) in a normal distribution. Thus, when the variance increases, the probability of getting small values of $x_i$ at $c_1$ or right after it is relatively high, which delays a shift detection. In the case of SRSD, the frequency distribution is more symmetric, because the probability of a potential change point passing the test increases toward the true change point $c_1$.

When the variance increases from one regime to another, ICSS performs particularly poor if the change point is located closer to the beginning of a time series. In an experiment with a change point at $c_1 = 25$, and a regime shift from one to three, ICSS found one change point in 46% of time, but these change points were all over the place, with only 1.1% of hits (Fig. 4c). In contrast, ICSS performed surprisingly well, being on par with SRSD, when a change point was shifted toward the end of a time series, unless it was too close to the end. Since SRSD is sequential method, designed to work in a monitoring mode, it is not surprising that it outperformed ICSS when a change point was placed at $c_1 = 90$ (Fig. 4d).

When the variance decreases from one regime to another, ICSS performs much better if a change point is shifted toward the beginning of a time series than toward the end. In an experiment when a change point was placed at $c_1 = 21$, ICSS and SRSD showed similar results (Fig. 4e), but when a change point was moved to $c_1 = 71$, the percentage of hits for SRSD was more than twice that of ICSS (Fig. 4f).

### 3.3 Series with two change points

In this group of experiments, the locations of change points remain constant at $c_j = (34, 67)$. The main goal here is to test different regime sequences. For the first regime sequence, $\sigma_j^2 = (1, 3, 1)$, ICSS correctly detected two change points in 39% of the series, versus 62% for SRSD (Table 4). In more than 50% of the series, ICSS detected no shifts at all. The accuracy of change point detection, as expressed by the prominence of the peaks at $c_1$ and $c_2$ in Fig. 5a is much lower for ICSS than for SRSD.

The situation becomes even worse for ICSS, if the lower variance regime is in the middle, i.e., $\sigma_j^2 = (3, 1, 3)$ (Fig. 5b). In this case, ICSS correctly detected two change points in only 21% of the series, and the accuracy of detection is about one third of that for SRSD. This is consistent with the results of Inclan & Tiao (1994), who found that ICSS performs best when the larger variance is in the middle, and its performance deteriorates in the opposite situation.

The most difficult situation, according to Inclan & Tiao (1994), is when the variances change in a monotone way, i.e., the variance increases at the first change point and increases again at the second change point. They estimate that, if this is the case, then it is necessary 500 or more observations for ICSS to be able to detect two change points in more than half of the time. Indeed, for a sequence of variances $\sigma_j^2 = (1, 3, 9)$ (Fig. 5c), ICSS detected two change points in only 44% of the series versus 69% for SRSD (Table 4). Especially problematic for ICSS was the detection of the first change point, the percentage of hits for which is much lower than for $c_2$. The detection statistics for $c_2$ is about the same for both methods.

When the variance decreased in a monotone way as $\sigma_j^2 = (9, 3, 1)$, ICSS detected the first change point more often than the second one, while SRSD detected both change points with about the same accuracy (Fig. 5d). The difference in the frequency of detection of two change points simultaneously between the two methods is quite substantial: 84% for SRSD and only 34% for ICSS (Table 5).

### 3.4 Series with outliers

Even a single outlier can have a substantial impact on regime shift detection procedure; it is like a monkey wrench thrown in the works. Figure 6 and accompanying Table 5 show a few examples of negative effects of outliers. In this set of experiments, the variance changes sixfold from one regime to another and an outlier ($x_i^*$) has a value of six. In the first example (Fig. 6a), a true change point was placed at $i = 34$ and an outlier at $i = 65$. Both SRSD and ICSS detected the change point equally well. The results for ICSS, however, revealed a spurious shift at $i = 65$. Interestingly, in 41% of the series ICSS found three or more change points. The outlier had practically no effect on SRSD.

The effect of an outlier was even more dramatic when it was placed closer to the true change point (Fig. 6b). In this case, ICSS detected one change point in 94% of the series, but the overwhelming majority of those change points were found at $i = 45$ (position of the outlier), not at $i = 34$ (true change point). Again, the results for SRSD were not affected, except for a small bump in the frequency distribution at $i = 45$ (Fig. 5b).

In some situations, an outlier may cause not a spurious shift, but rather a drastic deterioration of ICSS performance. For example, Figs. 5c and 5d show the results for the same variance regimes, $\sigma_j^2 = (6, 1, 6)$, and change points, $c_j = (35, 68)$. The only difference is that, in the latter case, there was an outlier at $i = 50$. As a result, both the sensitivity and accuracy of ICSS was drastically reduced. In the majority of the series (63%), ICSS found no change points at all, and the percentage of hits for $c_1$ and $c_2$ was reduced to 5.2 and 4.4, respectively. The performance of SRSD was not seriously affected by the outlier.

## 4 Examples of climatological time series

### 4.1 Arctic Ocean sea surface temperature

In recent decades, the Arctic is warming faster than other parts of the globe, a phenomenon known as Arctic Amplification. Some observational analyses found evidence for a 'wavier' jet stream in response to rapid Arctic warming (Francis and Vavrus, 2015), which, in turn, may lead to greater temperature variability in the Arctic. Figure 7a shows monthly sea surface temperature (SST) anomalies for the Arctic Ocean (65-90°N), based on the NOAA Optimum Interpolated SST dataset. This dataset (a.k.a. Reynolds OI.v2) is compiled using observations from ship inlets, buoys (both moored and drifting) and from satellites. Apart from an obvious warming trend, one can notice that SST fluctuations in recent years becomes more intense. The exact cause of increased SST variability in recent years are beyond the scope of this paper, but this time series is a good example to test the regime shift detection algorithms. First, it is necessary to detrend the SST series that can be done in a number of ways, depending on the researcher's view on what constitutes a trend. For this example, the LOWESS (local weighted regression) technique (Cleveland and Devlin, 1988) was used with a smoothing parameter of 0.1. When the LOWESS curve was subtracted from the SST anomalies, the residuals still contained a strong autocorrelation ("red noise"), therefore a prewhitening was needed. Given a lag-1 autocorrelation coefficient of 0.72 (OLS estimate), the prewhitening was performed as $e_i = x_i - 0.72 \, x_{i-1}$.

The result of the application of SRSD (0.05, 120, 3) to time series $\{e_i\}$ is presented in Fig. 7b. Three variance regime have been identified, with two change points: December 1996 and July 2007. The variance is lowest during the middle regime between these two change points ($s_2^2 = 0.003$). The variance is twice as high during the first regime ($s_1^2 = 0.006$) and almost three time as high during the most recent regime ($s_3^2 = 0.010$). These regime shifts are highly statistically significant, especially the second one, with the observed *p*-values of $1.2 \cdot 10^{-4}$ and $5.5 \cdot 10^{-9}$, respectively.

The ICSS algorithm found three change points, two of which coincided with those found by SRSD, and the third one was in September 2009. Formally, based on the F-test, the difference in the variances for two regimes, July 2007 – August 2009 (0.021) and September 2009 – November 2015 (0.007), is statistically significant at $p = 3.2 \cdot 10^{-4}$. The change point in September 2009 was not selected by SRSD for two reasons. First, the period July 2007 – August 2009 is only 27 months, that is, much shorter than the used cut-off length of 120 months. And second, given the Huber's tuning constant of 2, the weighted variance for this short regime is only 0.012, which makes the difference between the two regimes, statistically insignificant.

**4.2 Arctic Oscillation**

The Arctic Oscillation (AO) index is defined as the first empirical orthogonal function of sea level pressure in the Northern Hemisphere 20–90°N (Thompson and Wallace, 1998). Broadly speaking, positive values of the AO index indicate zonal atmospheric flow, while negative values suggest increased meridional circulation with ties to cold air advection into the midlatitudes (Thompson and Wallace, 2000). Analysing variability of the winter (DJF) AO index, Feldstein (2002) calculated that the ratio of its variance for the 1967–97 time period to that for the 1899–1967 period was 2.02 and concluded that it was well in excess of what to be expected if all of the interannual variability was due to atmospheric intraseasonal stochastic processes. Other studies (Hanna et al., 2015; Woollings et al., 2010) argue that AO variability is most likely a result of atmospheric internal variability, although it can be an early sign of destabilization of the polar jet stream and an increased susceptibility to external sources (Overland and Wang, 2015). Firm attribution of recent increased circulation variance is currently not possible (Trenberth et al., 2015).

Using the data since 1950, Overland and Wang (2015) noticed that the past decade showed the most variability in AO extremes for the month of December. Since they used the 9-yr running standard deviations, the timing of the transition to a high variance regime was rather vague. Figure 8a shows the result of the application of SRSD (0.1, 25, 2) to the time series of December AO index, 1950-2015. The change point in 2005 separates two regimes: 1) 1950-2004, with the variance of 0.91, and 2) 2005-2015, with the variance of 2.71. The latter includes the record and near-record high values in 2006 and 2011, respectively, as well as the record and near-record low values in 2009 and 2010, respectively. The difference in the variance for these two regimes is statistically significant at $p = 0.009$, based on the two-tailed F-test.

As noted in the previous section, if a change point is shifted toward the end of time series, ICSS performs better when the variance increases, then when it decreases. Despite that, ICSS failed to detect any change points in the December AO index, probably because the time series was too short for that method and the change point was too close to the end.

Both methods, however, were able to detect exactly the same change point in the monthly series of the AO index, January 1950 – December 2015 (Fig. 8b). The change point in April 2009 marks the transition from a low variance regime ($s_1^2 = 0.83$) to a high variance regime ($s_2^2 = 1.69$). The *p*-value for this regime shift is $1.1 \cdot 10^{-4}$.

**4.3 Temperature in the midlatitudes**

Analyzing long-term temperature records (since 1820) in eastern Minnesota, Skaggs et al. (1995) described a period of increased variability centered around 1920-1940, followed by sharp minimum in the middle to late 1960s, after which the variance started increasing again. They underscored, that this pattern of temperature variability was not characteristic of just eastern Minnesota. The period of low temperature variability in the late 1960s – early 1970s occurred more or less simultaneously over the United States. Using more recent gridded temperature data, Huntingford et al. (2013) examined changes to standard deviations before and after 1980. Although the choice of this change point year was made rather arbitrary, their map showed prominent changes over the midlatitudes for both hemispheres, with large parts of Europe and North America experiencing increased variability in more recent decades.

To illustrate the temporal pattern of changes in temperature variability noticed by Skaggs et al. (1995), Fig. 9 shows the first differences of mean winter (DJF) temperature in Minneapolis, MN, along with the 13-yr running standard deviations. Using the first differences is a simple way of eliminating a trend in the mean and focus on variability. The running standard deviations exhibit a general decline toward the middle of the series and an increase in more recent decades. Although running deviations are useful in depicting trends in variability, they make it hard to pinpoint the exact locations of regime shifts. Applying SRSD (0.1, 35, 3) to this time series revealed three temperature regimes with two change points, 1946 and 1980. During the first regime shift the variance decreased about 50%, from 1.59 to 0.74. The second regime shift was more prominent; the variance jumped threefold to 2.32. The *p*-values for the shifts were 0.02 and 0.001, respectively. Despite the high statistical significance of these shifts, ICSS failed to detect any of them.

A significant increase in summer temperature variability was also observed at numerous European station (Della-Marta et al., 2007; Parey et al., 2009; Yiou et al., 2009). For example, Della-Marta et al. (2007) found that over the period from 1880 to 2005 the length of summer heat waves over western Europe doubled and the frequency of hot days almost tripled. This trend of increasing temperature variability is projected to continue and even intensify in the 21st century (Fischer et al., 2012).

Figure 10 shows the results of a regime shift analysis for monthly surface air temperature (SAT) anomalies in western Europe (45-50°N, 0-10°E). This analysis was performed in two steps. First, regime shifts in the mean were identified using the sequential t-test algorithm included in SRSD (Rodionov, 2004). The tree detected regimes are presented in Fig. 10a as a stepwise trend. Although the differences between these regimes are seemingly small (their mean values are 0.16, -0.19, and 0.22, respectively), they are highly statistically significant, with the *p*-values of 0.003 for the shift down in February 1962, and $6.0 \cdot 10^{-6}$ for the shift up in August 1987. On the second step, this stepwise trend was removed from the original data, and the residuals were used as an input to SRSD and ICSS. The results for the two methods are exactly the same: three variance regimes with the change points between them in July 1971 and September 1981 (Fig. 10b). The *p*-values for the shifts in variance are 0.007 and $7.6 \cdot 10^{-5}$, respectively.

**4.3 El Niño – Southern Oscillation**

The magnitude of ENSO events has varied significantly over time, with multidecadal periods of strong and weak variability. Trenberth and Hoar (1996) noted strong ENSO variations from 1880 to the 1920s, and after about 1950 and, except for a strong event during 1939-1942, weaker variations from the mid-1920s to 1950. Exceptional ENSO variability can also be found in long ENSO-proxy records (Dunbar et al., 1994). Those periods usually coincided with large pre-industrial climate variations (Quinn and Neal, 1992). More recently, Hu et al. (2013) investigated an interdecadal shift in the variability and mean state of the tropical Pacific Ocean within the context of changes in ENSO. They concluded that, compared with 1979–99, the interannual variability in the tropical Pacific was significantly weaker in 2000–11.

The ability of SRSD and ICSS to detect periods of high and low variability of ENSO was tested on the Cold Tongue Index (CTI), one of the longest ENSO indices. The index is defined as the average SST anomaly over the region 6°N-6°S, 180-90°W minus the global mean SST. The index values are available from the University of Washington, Seattle, WA (http://research.jisao.washington.edu/enso/). According to the website, the global mean SST anomaly is subtracted from the CTI in order to remove a step shift upward at the onset of World War II when the composition of the marine observations largely changed from bucket to engine intake measurements and to lessen the secular trend in the time series that has been associated with global warming. Apparently, this has led to overcorrection of the CTI, because now there is a statistically significant ($p = 0.03$) step shift downward in 1943, as detected by SRSD (Fig. 11a).

Figure 11b shows the residuals (deviations from the regime means) and two change points detected by SRSD (0.05, 35, 3). These change points separate tree variance regimes: 1) High variance regime ($s_1^2 = 0.82$) in 1876-1919, 2) Low variance regime ($s_2^2 = 0.42$) in 1920-1982, and 3) High variance regime ($s_3^2 = 1.05$) that started with a powerful El Nino event of 1982/83. The *p*-values for the differences in variances between regime one and two is 0.01, and between regime two and three is 0.002. Despite the statistical significance of the shifts, ICSS failed to detect any of them.

**4 Summary and conclusions**

Changes in the variance of climatic time series are starting to attract more attention, especially from the perspective of global warming, because changes in the variance may have greater impact on temperature extremes than changes in the mean. In these circumstances, statistical methods capable of detecting shifts in the variance regimes are needed. A review of climate literature reveals that there are just a few practical methods available for that purpose and their applications are very limited so far. The situation is much more advanced in the area of econometrics, but the methods developed there need to be tested on climatic time series, which have their own specifics, such as a relatively short length and smaller magnitudes of shifts.

This paper compares two methods: SRSD developed by the author and ICSS developed by Inclan and Tiao (1994). The latter method is currently one of the most popular in econometric research. Both SRSD and ICSS are CUSUM-type methods, but while ICSS provides retrospective detection, SRSD utilizes a sequential approach, which makes it well suited for monitoring of regime shifts. Another difference is that, unlike ICSS, SRSD has two tuning parameters, the target probability level and cut-off length, that control the magnitude and time scale of variance regimes to be detected. This allows the users to tune up SRSD software, so that it better suits their needs. Also, it is important that SRSD has an internal mechanism for dealing with outliers. Due to the lack of such mechanism in ICSS, Inclan and Tiao (1994) advice to complement their procedure with some other procedure for outlier detection.

A comparison of the two methods was first implemented using Monte Carlo simulations with series of 100 points, which is a typical length (in years) of instrumental climatic time series. An interesting feature of ICSS is that it performs poor if a change point is located in the first half of a time series when the variance increases from the first regime to the second, or if a change point is in the second half of the series when the variance decreases. In contrast, ICSS performs better in the opposite types of situations, that is, when a change point is in the first half of a series in the case of decreasing variance, or it is in the second half in the case of increasing variance. With two change points, ICSS detects the second point much more often than the first one if the variance increases in a monotone way, that is, from the first regime to the second and then again to the third. If the variance decreases in a monotone way, ICSS more often selects the first change point than the second one. As for SRSD, it detects both change points with about the same degree of accuracy, regardless of the way the variance changes. Overall, SRSD outperforms ICSS in the overwhelming majority of the modelled situations. The only case when the two methods showed similar results was when a change point was located off the center of the series toward the end, but not too close to the end. For time series longer than 100 points, ICSS performs better, but it takes 500 or more points before it starts showing the results on par with SRSD. The experiments with series of more than 100 points, however, were very limited.

Some applications of ICSS for econometric time series have shown that it tends to overstate the number of actual change points in the variance (Fernández, 2004). This is probably only true for relatively long time series. As shown here, for time series of 100 points, ICSS rather tends to understate the actual number of change points, and in many cases fails to find any change points at all.

When time series contain outliers, SRSD has a clear advantage over ICSS. The effect of outliers on ICSS is quite interesting and may vary depending on both the absolute positions of outliers in the series and their positions in relation to the true change points. Most commonly, ICSS finds spurious change points at the positions of outliers, but just the presence of outliers can also make ICSS to grossly overstate the total number of change points in the series. The closer an outlier to a true change point, the more often it is mistakenly selected instead of a true change point. If an outlier is positioned between two change points, it may cause not a spurious shift, but rather a drastic deterioration of ICSS performance, when it cannot detect any shifts. The weighting method employed by SRSD helps effectively minimize the negative effects of outliers.

The methods were also evaluated by their performance on real climatic time series of different length. For longer time series with a monthly resolution (252 to 787 observations), ICSS and SRSD detected exactly the same change points in most cases. The only exception was the Arctic Ocean SST series, when ICSS found one extra change point that appeared to be spurious. For shorter time series with a yearly resolution (66 to 136 observations), ICSS failed to detect any change points, even when the variance doubled or tripled from one regime to another and the shifts were highly statistically significant. For this type of time series, SRSD is recommended.

It turned out that all the climatic time series considered in this paper, from the Arctic to the Tropics, has one thing in common: the last shift detected in each of these series was toward a high-variance regime. The increase in climate variability in recent decades and associated increase in frequency of extreme events have tangible impacts on society (Diaz and Murnane, 2008). Therefore, from this perspective and regardless of whether or not these changes are associated with global warming, documenting and monitoring regimes shifts in the variance are of primary importance.

**Table 1.** Percentages of total numbers of change points detected by SRSD ($p$, 30, 2) in each series on average and hits of the true change point $c_1 = 51$. The regime variances are $\sigma_j^2 = (1, 2)$.

| | Change points detected | | | |
|---|---|---|---|---|
| $p$ | 0 | 1 | $\geq 2$ | Hits |
| 0.05 | 57 | 41 | 2 | 4.5 |
| 0.1 | 41 | 54 | 5 | 5.9 |
| 0.3 | 13 | 65 | 22 | 7.7 |

**Table 2.** Percentages of total numbers of change points detected by SRSD (0.1, $l$, 2) in each series on average and hits of the true change points $c_j = (31, 61)$.

| $l$ | Change points detected | | | | Hits | |
|---|---|---|---|---|---|---|
| | 0 | 1 | 2 | $\geq 3$ | $c_1$ | $c_2$ |
| Regime variances $\sigma_j^2 = (1, 3, 1)$ | | | | | | |
| 20 | 12 | 13 | 70 | 5 | 12.6 | 14.5 |
| 30 | 13 | 14 | 73 | 0 | 17.7 | 15.7 |
| 40 | 46 | 38 | 16 | 0 | 6.8 | 13.3 |
| Regime variances $\sigma_j^2 = (3, 1, 3)$ | | | | | | |
| 20 | 2 | 9 | 69 | 20 | 15.8 | 15.8 |
| 30 | 3 | 17 | 79 | 1 | 16.9 | 16.5 |
| 40 | 21 | 69 | 10 | 0 | 15.0 | 15.0 |

**Table 3.** Percentages of total numbers of change points detected by SRSD ($p$, $l$, 2) and ICSS in each series on average in the case of one true change point. The corresponding frequency distributions of the detected change points are presented in Fig. 4.

| Fig. 4 | $c_1$ | $\sigma_j^2$ | $p$ | $l$ | SRSD | | | ICSS | | |
|---|---|---|---|---|---|---|---|---|---|---|
| | | | | | 0 | 1 | $\geq 2$ | 0 | 1 | $\geq 2$ |
| a | 51 | (1, 2) | 0.2 | 40 | 35 | 64 | 1 | 42 | 55 | 3 |
| b | 51 | (2, 1) | 0.2 | 40 | 20 | 79 | 1 | 41 | 57 | 2 |
| c | 25 | (1, 3) | 0.05 | 20 | 28 | 59 | 13 | 53 | 46 | 2 |
| d | 90 | (1, 3) | 0.1 | 20 | 67 | 19 | 14 | 51 | 47 | 2 |
| e | 21 | (3, 1) | 0.1 | 20 | 8 | 69 | 23 | 22 | 76 | 2 |
| f | 71 | (3, 1) | 0.1 | 25 | 15 | 71 | 14 | 28 | 69 | 3 |

**Table 4.** Percentages of total numbers of change points detected by SRSD ($p$, 25, $h$) and ICSS in each series on average in the case of two true change points $c_j = (34, 67)$. The corresponding frequency distributions of the detected change points are presented in Fig. 5.

| Fig. 5 | $\sigma_j^2$ | $p$ | $h$ | SRSD | | | | ICSS | | | |
|---|---|---|---|---|---|---|---|---|---|---|---|
| | | | | 0 | 1 | 2 | ≥3 | 0 | 1 | 2 | ≥3 |
| a | (1, 3, 1) | 0.05 | 2 | 23 | 15 | 62 | 0 | 55 | 5 | 39 | 1 |
| b | (3, 1, 3) | 0.05 | 2 | 17 | 27 | 56 | 0 | 62 | 16 | 21 | 1 |
| c | (1, 3, 9) | 0.2 | 3 | 0 | 28 | 69 | 3 | 0 | 54 | 44 | 2 |
| d | (9, 3, 1) | 0.2 | 3 | 0 | 9 | 84 | 7 | 0 | 64 | 34 | 2 |

**Table 5.** Percentages of total numbers of change points detected by SRSD (0.05, *l*, 2) and ICSS in each series on average in the case of experiments with an outlier $x_i^* = 6$. The corresponding frequency distributions of the detected change points are presented in Fig. 6.

| Fig. 6 | $c_j$ | $\sigma_j^2$ | Outlier | *l* | SRSD | | | | ICSS | | | |
|---|---|---|---|---|---|---|---|---|---|---|---|---|
| | | | | | 0 | 1 | 2 | ≥3 | 0 | 1 | 2 | ≥3 |
| a | 34 | (6, 1) | 65 | 25 | 0 | 88 | 12 | 0 | 1 | 54 | 4 | 41 |
| b | 34 | (6, 1) | 45 | 25 | 0 | 83 | 17 | 0 | 0 | 94 | 3 | 3 |
| c | (34, 68) | (6, 1, 6) | N/A | 20 | 0 | 4 | 90 | 6 | 31 | 4 | 61 | 4 |
| d | (34, 68) | (6, 1, 6) | 50 | 20 | 2 | 10 | 85 | 3 | 63 | 21 | 14 | 2 |

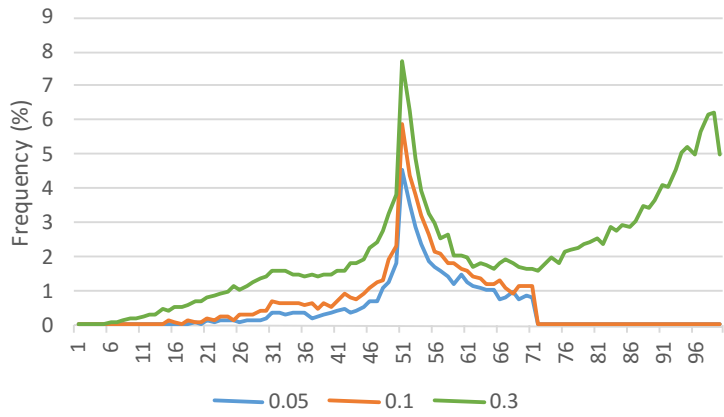

**Figure 1.** Frequency distribution of change points detected by SRSD ($p$, 30, 2) for three different values of $p$: 0.05, 0.1, and 0.3. The true change point is $c_1 = 51$ and regime variances are $\sigma_j^2 = (1, 2)$.

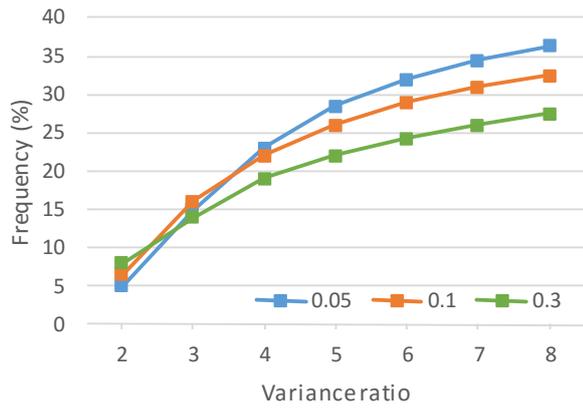

**Figure 2**. Percentage of hits as a function of the variance ratio between two regimes ($\sigma_2^2/\sigma_1^2$) for different values of $p$ in SRSD ($p$, 30, 2).

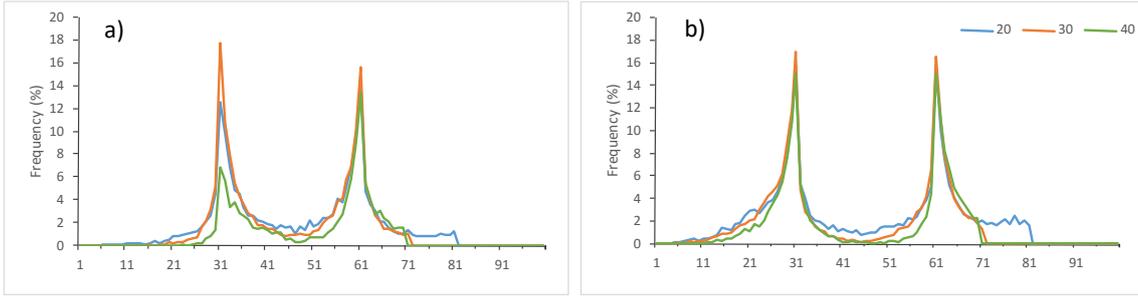

**Figure 3**. Frequency distributions of change points detected by SRSD (0.1, $l$, 2) with three different values of $l$: 20, 30, and 40. The true change points are $c_j = (31, 61)$, and regime variances are a) $\sigma_j^2 = (1, 3, 1)$ and b) $\sigma_j^2 = (3, 1, 3)$.

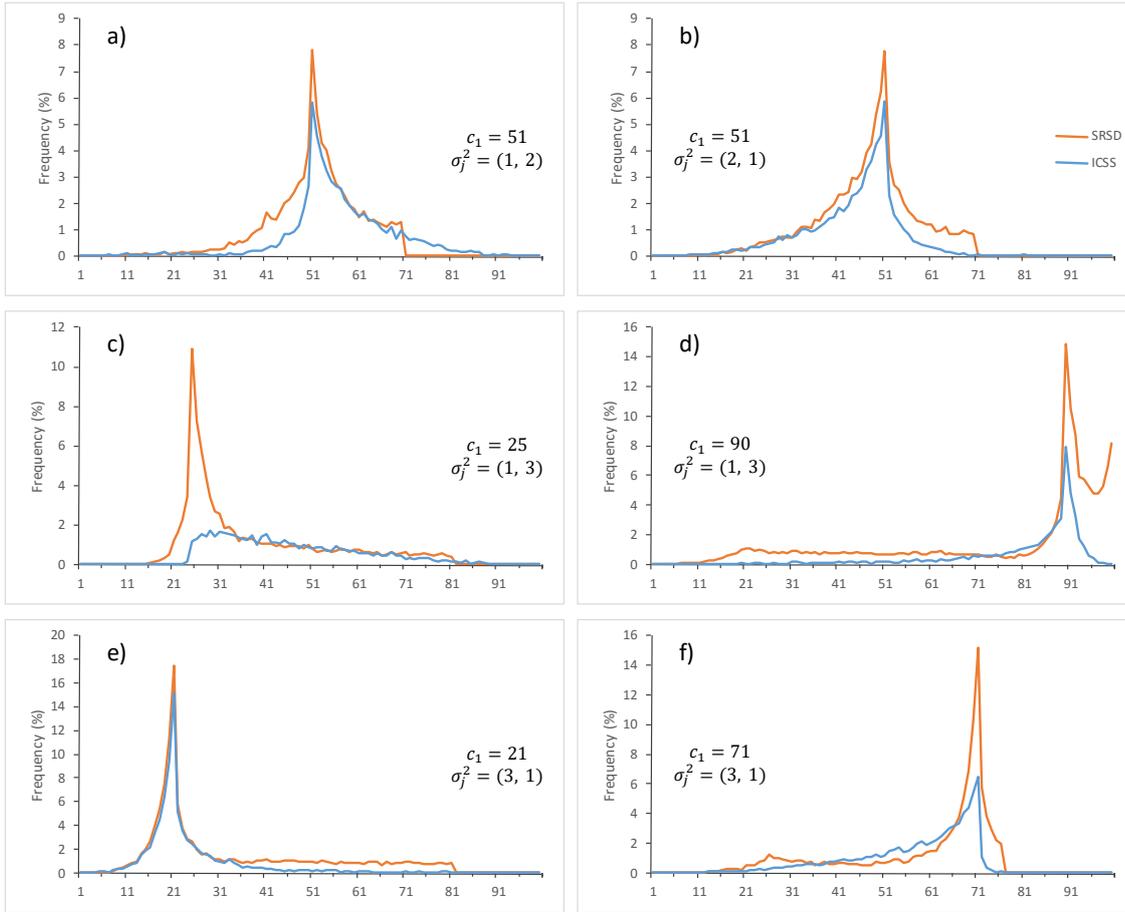

**Figure 4.** Results of Monte Carlo experiments with one change point. The parameters of SRSD used in each experiments are given in Table 3.

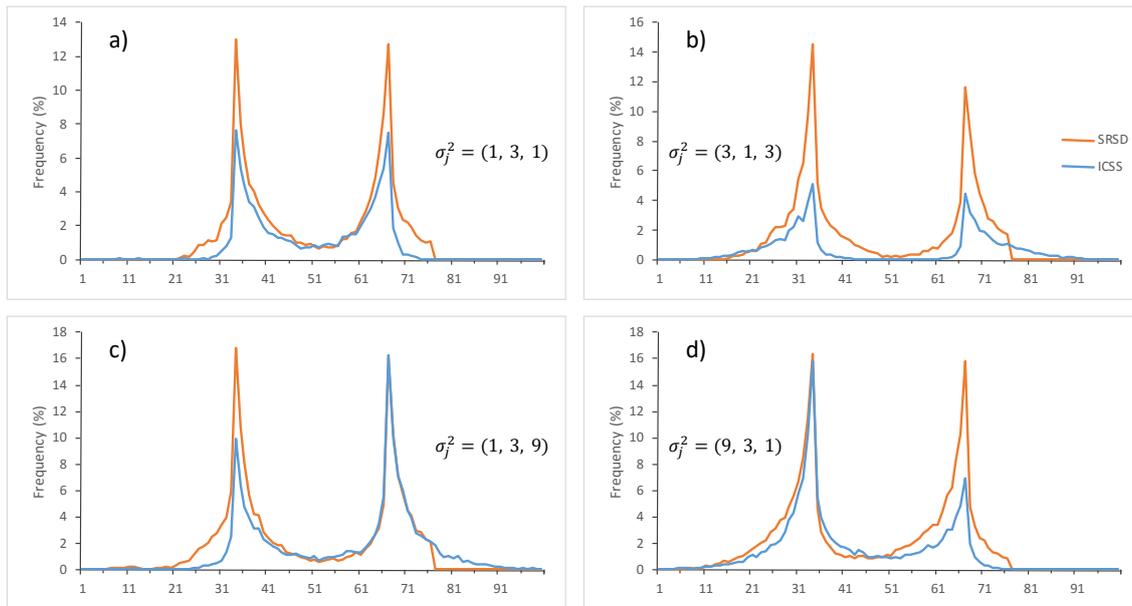

**Figure 5.** Results of Monte Carlo experiments with two change points $c_j = (34, 67)$. The parameters of SRSD used in each experiments are given in Table 4.

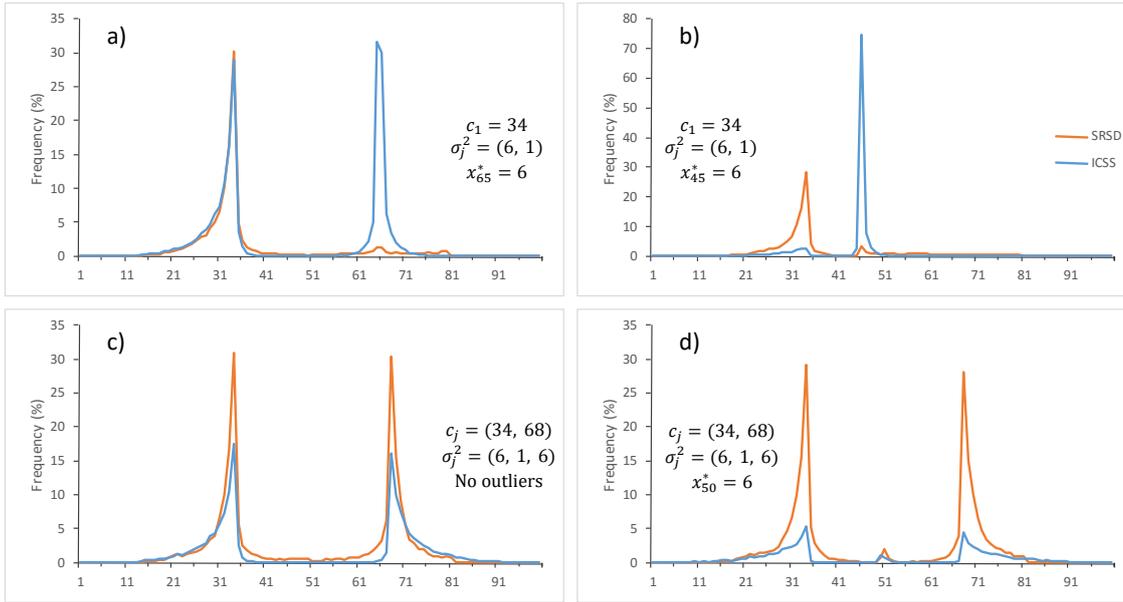

**Figure 6.** Results of Monte Carlo experiments with an outlier $x_i^* = 6$. The parameters of SRSD used in each experiments are given in Table 5.

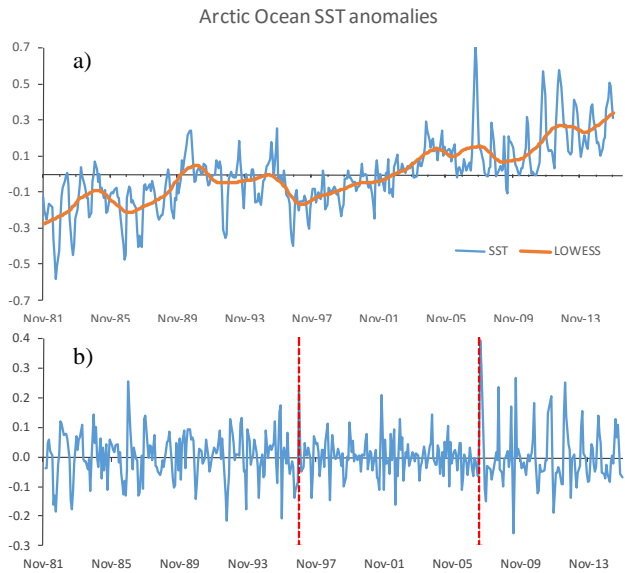

**Figure 7.** Arctic Ocean SST, November 1981 – December 2015: a) SST anomalies (deviations from the period mean normalized by standard deviations for each month) and the LOWESS curve with a smoothing parameter of 0.1; b) residuals (after removing the LOWESS curve and prewhitening) with two change points, December 1996 and July 2007, detected by SRSD (0.05, 120, 3). ICSS detected the same change points plus a third change point in September 2009 with a shift toward lower variance.

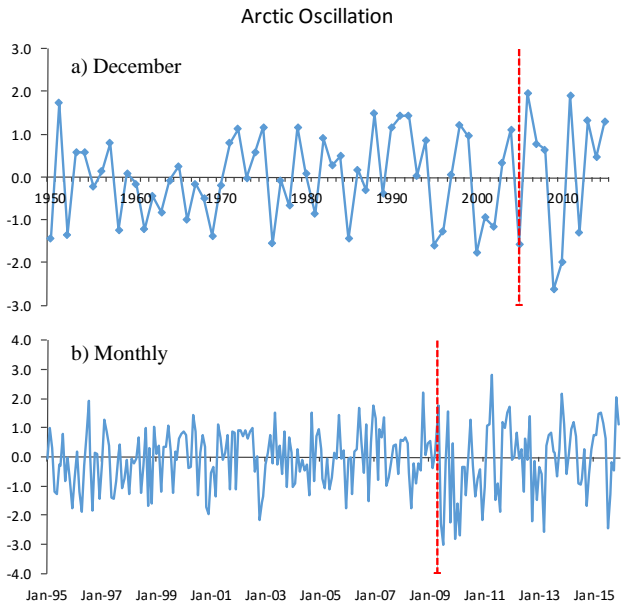

**Figure 8.** The Arctic Oscillation index: a) December values, 1950 – 2015, with a change point in 2005 detected by SRSD (0.1, 25, 2); b) monthly values, January 1995 – December 2015, with a change point in April 2009 detected by SRSD (0.05, 120, 2) and ICSS.

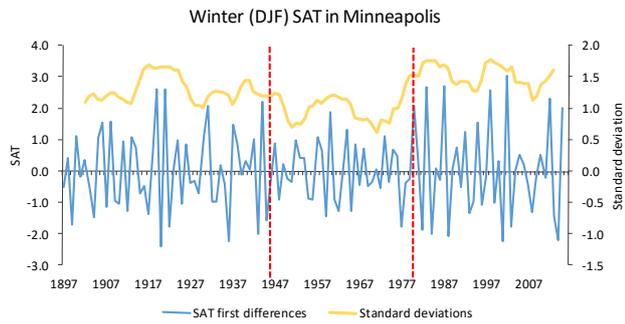

**Figure 9.** First differences of winter (DJF) SAT anomalies in Minneapolis, MN, along with the 13-yr running standard deviations and two change points (1946 and 1980) detected by SRSD (0.1, 35, 3).

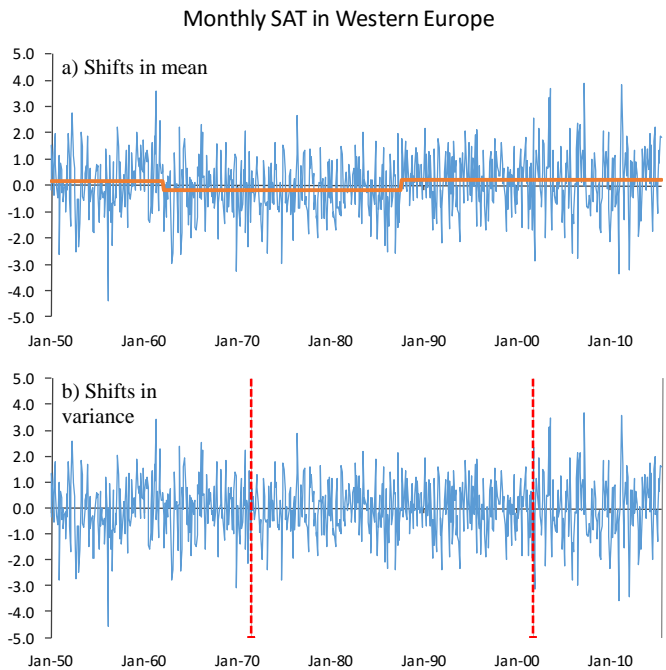

**Figure 10.** a) Normalized (by standard deviations) monthly surface air temperature (SAT) anomalies in western Europe (45-50°N, 0-10°E) and regime shifts in the mean detected by SRSD (0.05, 120, 2). The change points are February 1962 and August 1987; b) Residuals after removing the stepwise trend in the mean and regime shifts in the variance detected by SRSD (0.1, 120, 2) and ICSS. The change points are July 1971 and September 2001.

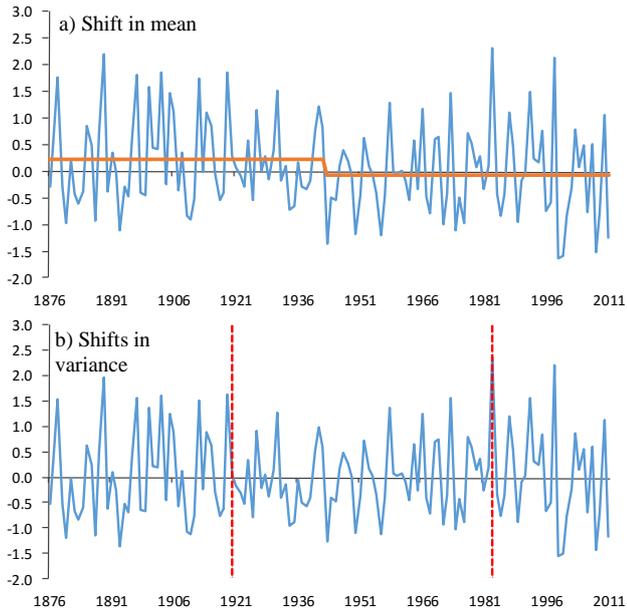

**Figure 11.** a) Mean winter (DJF) Cold Tongue Index, 1876 – 2011, showing a step shift downward in 1943 as detected by SRSD (0.1, 35, 3); b) Residuals after removing the stepwise trend in the mean and regime shifts in the variance detected by SRSD (0.05, 35, 3) with change points in 1920 and 1983.